\documentclass{article}
\usepackage{slashed,epsfig,amsmath,amssymb,enumitem} \usepackage[papersize={8.5in,11in}]{geometry}
   \usepackage{bm}
\usepackage{graphicx}
\newcommand{\be}{\begin{equation}}
\newcommand{\ee}{\end{equation}}
\title{Void and Density Walls Inhomogeneous Cosmic Web and Dark Energy}
\author{J. W. Moffat\\
Perimeter Institute for Theoretical Physics, Waterloo, Ontario N2L 2Y5, Canada\\
and\\
Department of Physics and Astronomy, University of Waterloo, Waterloo,\\
Ontario N2L 3G1, Canada}

\begin{document}

\maketitle


\begin{abstract}
 Recent observations from the Dark Energy Spectroscopic Instrument (DESI 2025) indicate a weakening of cosmic acceleration at low redshifts $z < 1$, with effective dark energy equation of state parameters $w_0 > -1$ and $w_a < 0$. We demonstrate that this evolution in dark energy can be explained by cosmic inhomogeneities and the Cosmic Web without modifying fundamental physics. Our model shows how the differential expansion between underdense voids and overdense walls creates an effective backreaction term that simulates evolving dark energy when interpreted within homogeneous cosmological frameworks. The inhomogeneous cosmic structure formation becomes significant $z\sim  1-2$, the increasing gravitational influence of wall regions counteracting the cosmic acceleration, producing both a weakening acceleration signal and a direction-dependent Hubble parameter consistent with local measurements. This mechanism reconciles the higher locally measured Hubble constant $H_0\approx 73~ km/s/Mpc$ with the lower value inferred from CMB observations $H_0\approx  67-69~ km/s/Mpc$ without introducing new energy components or modifying general relativity. Our model makes testable predictions regarding directional and scale-dependent variations in cosmological parameters that can be verified with next-generation surveys. This work suggests that properly accounting for cosmic structure may be essential for resolving apparent tensions in cosmological parameters.
\end{abstract}

\section{Introduction}

The standard model of cosmology, Lambda Cold Dark Matter $\Lambda$CDM, has been remarkably successful in explaining a wide range of cosmological observations, from the cosmic microwave background (CMB) anisotropies to the large-scale structure of the Universe. However, as measurement precision has improved, significant tensions have emerged that challenge the internal consistency of this framework. The most prominent of these is the Hubble tension, a statistically significant discrepancy between the value of the Hubble constant $H_0$ measured from local distance indicators and that inferred from CMB observations assuming the $\Lambda$CDM model~\cite{Riess2024,PlanckCollaboration2020}.

Local measurements using Cepheid-calibrated Type Ia supernovae consistently yield 
$H_0\approx  73~ km/s/Mpc$~\cite{Riess2024}, while CMB-based determinations from Planck give $H_0\approx 67-68~ km/s/Mpc$~\cite{PlanckCollaboration2020}. This $9\%$ difference now exceeds $5\sigma$ in statistical significance, making it increasingly difficult to attribute to systematic errors in either measurement technique. The persistence and significance of this tension have motivated numerous theoretical proposals for its resolution, ranging from modified dark energy models to new physics in the early Universe~\cite{DiValentino2021}.

Adding to this cosmological puzzle, recent results from the Dark Energy Spectroscopic Instrument (DESI 2025) have revealed evidence for a weakening of cosmic acceleration at low redshifts $z < 1$. These observations suggest an effective dark energy equation of state parameter $w_0 < -1$, deviating significantly from the cosmological constant value of $w_0 = -1$ predicted by the standard $\Lambda$CDM model. This apparent evolution of dark energy presents both a challenge to the standard model and a potential clue toward resolving existing tensions. Large-scale structure (LSS) surveys have revealed that the distribution of matter in the Universe is far from homogeneous and isotropic on the scales of galaxies and galaxy clusters. Observations indicate a Cosmic Web of voids—underdense regions and walls-filaments structures—partitioning space into an inhomogeneous environment. A void-wall cosmological model posits that the differential expansion rates of these underdense and overdense regions can, on average, slow down an effective acceleration that no longer requires a strict cosmological constant. The inhomogeneities can act as a source term in the averaged Friedmann equations, leading to a form of effective dark energy.

We propose that both the Hubble tension and the apparent evolution of dark energy can be explained within the framework of void-wall cosmology, without invoking new fundamental physics~\cite{Moffat}. Our approach focuses on the effects of cosmic structure, specifically, the influence of inhomogeneities on the average cosmic expansion. As the Universe evolves and structure formation progresses, the differential expansion between underdense voids and overdense walls creates an effective backreaction term that, when interpreted within homogeneous cosmological frameworks, appears as evolving dark energy. The timing of significant structure formation $z\sim  1-2$ naturally coincides with the observed onset of weakening cosmic acceleration. As matter clusters into walls, their stronger gravitational attraction counteracts the repulsive effect of dark energy, effectively weakening the overall cosmic acceleration. This mechanism not only explains the DESI 2025 observations but also provides a pathway to resolving the Hubble tension through scale-dependent and direction-dependent effects on the measured expansion rate.

Our approach builds upon previous work on cosmic inhomogeneities ~\cite{Buchert1,Buchert2,Buchert3}, but extends it to make specific, testable predictions regarding the relationship between cosmic structure and apparent dark energy evolution. We demonstrate through analytical calculations that the magnitude of these effects is sufficient to account for the observed Hubble tension, and the DESI 2025 evidence for a weakening cosmic acceleration and dark energy, while remaining consistent with other cosmological constraints.

The purpose of this paper is to outline the basic theoretical foundations of the void-wall cosmic web inhomogeneous model and how it yields an effective evolving dark energy with $w_0\neq -1$. The predictions of the void-wall framework are applied to the DESI data~\cite{DESI20251,DESI20252,DESI20253,DESI20254}, which when combined with additional datasets, suggest a possible departure from $\Lambda$CDM at the level of dark energy dynamics. Through an analysis of separate scale factors for voids and walls, volume partitioning, and backreaction terms in the averaged Friedmann equations, we demonstrate that the void-wall approach accommodates a dynamical effective dark energy component. Comprehensive numerical modeling and direct confrontation with observational data are required to validate and refine this scenario, but even at the qualitative level, it provides a compelling alternative to the conventional assumption of a constant $\Lambda$ as the sole driver of cosmic acceleration.

The paper is organized as follows: Section 2, describes the physical components and evolution of the Cosmic Void. In Section 3, the void-wall inhomogeneous model is developed and reviews the observational evidence for both the Hubble tension and evolving dark energy. Section 4, presents a resolution of the Hubble tension and the derivation of the late time weakening of the dark energy an develops testable predictions of our model. Finally, Section 5 summarizes our conclusions and outlines directions for future work.

\section{The Cosmic Web}

The Cosmic Web represents the largest known structure in the Universe, a complex network characterized by four primary components that together form the scaffolding of the cosmos~\cite{Colless,Tegmark}. Vast underdense void regions occupying around $80\%$ of the Universe's volume but containing only around $20\%$ of its mass. The typical void size is 30-150 Mpc in diameter corresponding to 100-500 million light-years. The void density is $10-30\%$ of the cosmic average density. The void internal structure is not completely empty but containing tenuous gas and occasional isolated galaxies. The voids expand faster than the cosmic average, contributing to the apparent acceleration.

Physical sheets are flattened overdense structures forming the boundaries or walls between voids with a typical thickness 5-10 Mpc  or 15-30 million light-years. They can span around 100 Mpc across with a density 1-5 times the cosmic average density. They contain numerous galaxies arranged in flattened distributions that expand slower than the cosmic average due to self-gravity

Cosmic filaments consist of elongated, thread-like structures connecting walls and nodes with typical diameter 2-5 Mpc or 6-15 million light-years and a typical length 50-100 Mpc or 150-300 million light-years. Their density is 5-20 times the cosmic average density. They contain galaxies arranged in linear patterns and significant amounts of warm-hot intergalactic medium (WHIM). The filaments exhibit complex flow patterns with matter streaming along their length toward nodes.

Cosmic nodes are clusters of dense concentrations at filament intersections of typical size 2-5 Mpc in diameter and a density 50-500 times the cosmic average density.
They host massive galaxy clusters containing thousands of galaxies and contain extremely hot intracluster gas $10^7 - 10^8$ K. They are often not in equilibrium, with ongoing mergers and accretion.

The Cosmic Web structure is primarily shaped by unidentified $84\%$ dark matter content and visible matter galaxies. Gas traces the underlying dark matter distribution and $30-40\%$ of all baryons reside in the warm-hot intergalactic medium $10^6-10^7$ K within filaments. The large-scale tidal field of the Universe plays a crucial role in shaping the Cosmic Web. Structure formation proceeds through anisotropic gravitational collapse, forming sheets first, then filaments, and finally nodes. The coherent velocity flows exist along filaments toward nodes with typical speeds of $200-500~ km/s$

The Cosmic Web's development follows a clear temporal progression. Early Cosmic Web seeds form at $z > 100$ with initial density fluctuations from quantum fluctuations in the early Universe. It is largely homogeneous with only small statistical fluctuations.
The first Cosmic Web structures begin $z < 100$ and the initial overdensities begin growing linearly with no visually distinct Cosmic Web yet. The emergence of the Web occurs at a redshift 
$5 < z < 20$ and first recognizable dense sheets begin forming.
Hierarchical clustering begins and reionization occurs as first stars form in early nodes. At a redshift $2 < z < 5$ distinct filamentary structure becomes clearly visible
and the first massive galaxy clusters form at nodes. Voids begin to empty as matter flows toward filaments. At the maturation phase $0.5 < z < 2$ the Cosmic Web reaches recognizable modern configuration, significant void emptying occurs and wall and filament structure becomes well-defined. The phase coincides with peak cosmic star formation. The current phase  at redshift $z < 0.5$ displays continued evolution with increasing contrast between structures
        
The period around $z\sim 1-2$ represents a critical transition in Cosmic Web evolution.
The web transitions from formation to maturation with significant increase in density contrast between voids and filaments and walls. The gravitational effects of the web structure become cosmologically significant and coincide with dark energy beginning to dominate the cosmic expansion.

The Cosmic Web is observed through multiple complementary techniques. The map of the 3D distribution of galaxies e.g., SDSS, DESI detects the dark matter distribution directly and the Lyman-alpha Forest traces neutral hydrogen in filaments. X-ray observations detect hot gas in clusters and filaments. The Sunyaev-Zeldovich effect maps hot gas through CMB photon scattering and the Fast Radio Bursts probe ionized gas in the Cosmic Web through dispersion measure. The Cosmic Web represents not just the largest structure in the Universe but a comprehensive record of cosmic evolution, extending from the present day back to $z\sim 20$ in directly observable form, with its seeds tracing back to the earliest moments after the Big Bang.

\section{Void-Wall model}

In a void-wall model, underdense voids and overdense walls or filaments can each be approximated by their own local scale factors $a_v(t)$ and $a_w(t)$, which evolve differently over cosmic time t. Voids typically expand faster than walls because of their lower matter density. An ansatz is to split the total volume of the Universe into void and wall components. Let the fraction of the total cosmic volume occupied by voids be $f_v(t)$ and by walls $f_w(t)$. These fractions evolve over time because voids generally grow at the expense of walls as structures form and become more clumped.
An overall or averaged Hubble parameter $H_{\rm eff}(t)$ can be defined by combining the separate Hubble parameters of voids $H_v(t)=\dot a_v(t)/a_v(t)$ and walls $H_w(t)=\dot a_w(t)/a_w(t)$ where $\dot a(t)=da(t)/dt$, weighted by their fractional volumes:
\be
H_{\rm eff}(t)=f_v(t)H_v(t)+f_w(t)H_w(t).
\ee
If $H_w(t) > H_v(t)$ and $H_w(t)$ occupies a larger fraction $f_w(t)$ of the volume at late times, the total effective expansion rate $H_{\rm eff}(t)$ can decrease relative to a homogeneous model, thus impacting measured values of the Hubble parameter at low redshift.

In inhomogeneous scenarios, the global expansion rate can deviate from the homogeneous constant $\Lambda$CDM expectation. When observational teams like the Dark Energy Spectroscopic Instrument (DESI) fit the cosmic expansion history to a parameterized dark energy equation of state~\cite{Linder1,Linder2}:
\be
w(z) = w_0+(1-a(z))w_a=w_a\left(\frac{z}{1+z}\right),
\ee
where $w_0$ denotes the value of w today and a(z) is the cosmological scale factor. As displayed in Fig. 1, recent DESI data indicate that $w_0 > -1$ and $W_a < 0$~\cite{DESI20251,DESI20252,DESI20253,DESI20254}. This raises the possibility that structure formation and void and wall growth may influence the apparent late-time acceleration. The volume fraction of voids decreases due to the increasing collapse of the density walls over time, amplifying the slower local expansion rate in the overderdense wall regions and modifying the measured H(z) compared to a homogeneous model with identical parameters. As a result, the inhomogeneous expansion can appear as a time-varying dark energy component.

In some treatments often following Buchert’s averaging formalism~\cite{Buchert1,Buchert2,Buchert3}, the inhomogeneities introduce additional terms in the averaged Friedmann equations. One can formally write:
\be
\langle H^2\rangle=\frac{8\pi G\langle\rho\rangle}{3} +\frac{\Lambda}{3}-\frac{1}{6}\langle Q\rangle- \frac{k_D}{a_D^2},
\ee
where $\langle\rho\rangle$ is the average matter density, $\langle Q\rangle$ encodes the  average effect of inhomogeneities, $k_D$ the domain average curvature and $a_D$ is the domain-averaged scale factor. 

The statement that the cosmological data finds that the Universe is spatially flat with k=0 refers to measurements in the standard FLRW cosmology where homogeneity and isotropy are assumed. 
The term $k_D/a_D^2$ represents the domain-averaged spatial curvature contribution. This is distinct from the global curvature parameter k in the standard FLRW model. The implication of observational evidence for a spatially flat Universe with k=0 is that the observed flatness may be an effective flatness rather than a true absence of curvature in the inhomogeneous universe.
It is possible that $k_D/a_D^2$ is non-zero, but its effects are compensated by the backreaction term $\langle Q\rangle$ in such a way that observations interpret the universe as flat. The domain-averaged curvature $k_D$ could be scale-dependent, meaning it might vary depending on the averaging domain size. The measured k=0 in standard $\Lambda$CDM cosmology does not necessarily imply that $k_D/a_D^2=0$ in the averaging formalism. Instead, what we observe as spatial flatness could be the combined effect of actual domain curvature and backreaction from inhomogeneities.
This is one of the interesting aspects of the averaging formalism; it allows for the possibility that what we measure as dark energy could partially be a consequence of averaging inhomogeneities, with the actual geometry of space potentially being more complex than the simple flat case

One can interpret $\langle Q\rangle$ by moving it to the matter side of the Einstein field equations and treating it as an effective fluid with energy density 
$\rho_{\rm eff}$ and pressure $p_{\rm eff}$. This effective fluid can yield:
\be
w_\mathrm{eff}
\;\equiv\;
\frac{p_\mathrm{eff}}{\rho_\mathrm{eff}}
\;\neq\;
-1,
\ee
and can vary with redshift z. The difference from -1 reflects a dynamical departure from a pure cosmological constant $\Lambda$.

The inhomogeneity void-wall backreaction can accumulate over cosmic time, even from comparatively small local density contrasts, and simulate or alter a dark-energy–like component at low redshift. This does not require that every local void or wall be extremely deep or massive; rather, the integrated average effect across the cosmic web can yield a shift in late-time cosmic expansion relative to homogeneous models.

Because of the volume partitioning in a void-wall model, the effective expansion can fit observational data with $w_0\neq -1$ and $w_a\neq 0$. Growing wall fraction with time drives a differential expansion that can look like a slowing down time-dependent dark energy. The large-scale structure inhomogeneities shift the measured expansion history away from the baseline $\Lambda$CDM extrapolation expectation. To see how $w_0\neq -1$ can arise, consider schematically:
\be
H_\mathrm{eff}^2(z)
\;=\;
H_0^2\,
\Bigl[\,
\Omega_m\,f_m(z)
\;+\;
\Omega_\Lambda
\;+\;
\Omega_\mathrm{inh}(z)
\Bigr],
\ee
where $\Omega_{\rm inh}(z)$ is the contribution from the void-wall inhomogeneities that can depart from a strict constant. One can rewrite $\Omega_{\rm inh}(z)$ in terms of an effective dark energy density $\rho_{\rm DE,eff}$ with equation of state $w_{\rm eff}$. The time or redshift dependence of $\rho_{\rm DE,eff}$ can make $w_{\rm eff}\neq -1$.

The total effective dark energy contribution would be:
\be
\Omega_{\text{DE,eff}}(z) = \Omega_\Lambda + \Omega_{\text{inh}}(z).
\ee
For dark energy to weaken as z decreases as we approach the present time, we need the effective equation of state $w_{\rm eff} > -1$, which means the dark energy density dilutes faster than a cosmological constant. This can happen if $\Omega_{\rm inh}(z)$ has a negative contribution that grows in magnitude as z decreases. In the void-wall model, as structure formation proceeds, the universe becomes increasingly partitioned into underdense voids and overdense walls and filaments.
If the wall fraction grows with time and decreasing z, this differential expansion can create a backreaction effect that partially counteracts the acceleration from $\Omega_\Lambda$.
The $\Omega_{\rm inh}(z)$ becomes more negative as z decreases, effectively reducing the total dark energy contribution.

This is how large-scale structure inhomogeneities could shift the measured expansion history away from the baseline $\Lambda$CDM expectation, simulating a time-dependent dark energy component that appears to be slowing down over cosmic time. An explicit example is given by
\be
\Omega_{\rm inh}(z)=\Omega_Q(z)-\Omega_{kD},
\ee
where $\Omega_Q$ is associated with the inhomogeneity, and $\Omega_{kD}$ is the averaged curvature term. If $\Omega_Q$ evolves nontrivially with redshift, that leads to an effective $w_{\rm eff}\neq -1$. Surveys like DESI, by measuring the expansion rate H(z) and growth of structure at multiple redshifts, can detect or constrain this departure.

As the Universe evolves, voids grow in comoving size and volume fraction. Their lower matter content causes them to expand more rapidly, effectively boosting the global expansion rate at late times. Overdensity walls hold on to matter more strongly, limiting local expansion, and if they dominate enough volume at late times keep $H_{\rm eff}$ from rising. The spatial gradients and differential expansion rates feed into the backreaction terms, shifting the effective Friedmann evolution in a way that can look like evolving dark energy.  At late times, one can schematically split the total energy budget into two components, the usual cosmological constant term and the contribution from inhomogeneities:
\be
\Omega(z)=\Omega_\Lambda+\Omega_{\rm inh}(z).
\ee
The void-wall Cosmic Web cosmology model treats $\Omega_\Lambda$ as a constant cosmological constant vacuum energy in the background model, while $\Omega_{\rm inh}$ encodes the effects of the void-wall inhomogeneous Cosmic Web. In the high redshift early Universe, the void-wall contrast is negligible, so for $ z >> 1$, $\Omega_{\rm inh}\rightarrow 0$. In this limit, one recovers the standard $\Lambda$CDM form with $\Omega_\Lambda\sim 0.7$ and the matter, radiation dominated era,  fitting the angular CMB power spectrum and other early-Universe observables. At late times, as overderdense walls grow, their slower expansion enhances $\Omega_{\rm inh}$. The slower accelerated expansion then appears partly attributable to $\Omega_\Lambda$ and partly to the inhomogeneity-driven term $\Omega_{\rm inh}$. Because $\Omega_{\rm inh}(z)$ can vary with redshift, standard $\Lambda$CDM fitting, which assumes purely constant $\Omega_\Lambda$ may infer an effective dark energy equation of state parameter $w_0\neq -1$.  

In essence, the inhomogeneous Cosmic Web leaves the early-Universe physics—dominated by $\Omega_\Lambda$ and other $\Lambda$CDM-like parameters—nearly unaltered, while generating a significant extra contribution at lower redshifts. This extra term can modify the late-time acceleration. Consequently, the void-wall picture still reproduces CMB constraints yet allows for departures from a strict cosmological constant in the late Universe.

In the standard treatment of cosmology, including both $\Lambda$CDM and the inhomogeneous void-wall cosmic web scenario, the transition from decelerating to accelerating expansion occurs at a redshift of order $z\sim 0.5 - 1$ at around 5 billion years after the Big Bang. In the conventional homogeneous $\Lambda$CDM model, the dark energy begins to dominate over matter near this epoch, causing the onset of cosmic acceleration. In a void-wall inhomogeneous model, this same redshift range marks the era when walls and overderdense regions begin to occupy a larger fraction of the total cosmic volume, decreasing the net expansion rate as compared to a purely homogeneous Universe. In the void-wall cosmology, the effective dark energy onset of acceleration generally coincides with the growth of the inhomogeneous Cosmic Web. At higher redshifts, $z > 2$, matter is still relatively uniform on large scales, meaning voids and walls have not yet reached sufficient contrast for their differential expansion to significantly affect the overall Hubble rate. By the time we reach redshifts $z\leq 1$, walls are expanding slower, voids remain more compressed, and the associated backreaction or volume-weighted averaging can manifest as an effective late-time slow down of acceleration similar to dark energy with an evolving equation of state.

At very high redshift, the underdense and overdense regions have not yet become sufficiently pronounced to alter the overall expansion. From the CMB’s perspective, the Universe effectively behaves very much like a homogeneous fluid dominated successively by radiation and then matter, so fitting early-time data $\Lambda$CDM with a small or negligible dark energy fraction. In a void-wall framework, the separate scale factors for voids and walls $a_v$ and $a_w$ are nearly identical during this era, and any inhomogeneity-induced effective fluid is negligible. As time goes on and the redshift approaches $z\sim 1$, walls grow in volume fraction $f_w$. Since they are overderdense, their expansion rate $H_w$ is slower than that of voids $H_v$. The total averaged Hubble parameter
\be
H_{\rm eff}=f_vH_v+(1-f_v)H_w,
\ee
starts to decrease below a uniform matter-dominated Universe would predict. An observer fitting a homogeneous cosmology to this late-time expansion sees a smaller-than-expected acceleration. This effect can appear as a fluid with $w_0\neq -1$, because it grows in importance only after structure formation has given rise to significant density contrasts. In standard $\Lambda$CDM, we assume a constant dark energy density $\rho_\Lambda$. Since the inhomogeneities are negligible in the early Universe, this approach becomes a $\Lambda$CDM-like evolution for CMB-era data. One can still fit the CMB well if the Cosmic Web inhomogeneity is effectively turned off until lower redshifts. Around $z\sim 0.5- -1$, the growth of voids and walls begins to yield a dark energy signal, but it need not have $w_0 = -1$. Observers measuring the expansion rate at these redshifts can thus infer a dynamical w(z).  

The cosmological constant-like behavior at high redshift blends seamlessly with a dynamically evolving dark energy at lower redshift. Rather than requiring an actual fluid with a constant vacuum energy for all epochs, the void-wall framework treats backreaction from inhomogeneities as negligible at early times, thereby fitting the CMB and significant at late times explaining acceleration with $w_0\neq -1$.

When DESI or other experiments fit observations to a parameterized w(z), they infer $w_0\neq -1$ and $w_a\neq 0$, and fundamental scalar fields or modified gravity models are invoked to explain the dynamical behavior of the dark energy. Instead, the observed acceleration or time-variation in the equation of state reflects the complex geometry of voids and walls.

Because a void-wall cosmology includes (i) different local scale factors, (ii) volume fractions that evolve over time, and (iii) a nontrivial backreaction, the global expansion can appear as a dynamical dark energy with $w_0\neq -1$, matching DESI findings. This effectively ties the late-time acceleration and equation-of-state variation to inhomogeneous structure formation, rather than requiring a purely constant $\Lambda$. Validating further such a scenario requires detailed numerical work, solving the coupled void-wall evolution equations, fitting large-scale-structure data, and comparing the inferred H(z) and $\sigma_8$ to observations. Nonetheless, the key result is that inhomogeneous expansion provides a natural mechanism for what appears, on average, as a time-dependent dark energy component.

\section{Time-dependent dark energy and the Hubble tension}

In dynamical dark energy models, the dark energy density evolves as:
\be
\rho_{DE}(z) = \rho_{DE,0} \exp\left(3\int_0^z \frac{1+w(z')}{1+z'} dz'\right),
\ee
where w(z) is the equation of state parameter that can vary with redshift z.
This changes the Hubble parameter evolution:
\be
H^2(z) = H_0^2 \left[\Omega_m(1+z)^3 + \Omega_r(1+z)^4 + \Omega_{DE}\frac{\rho_{DE}(z)}{\rho_{DE,0}}\right].
\ee

\begin{figure}
    \centering
    \includegraphics[width=0.75\linewidth]{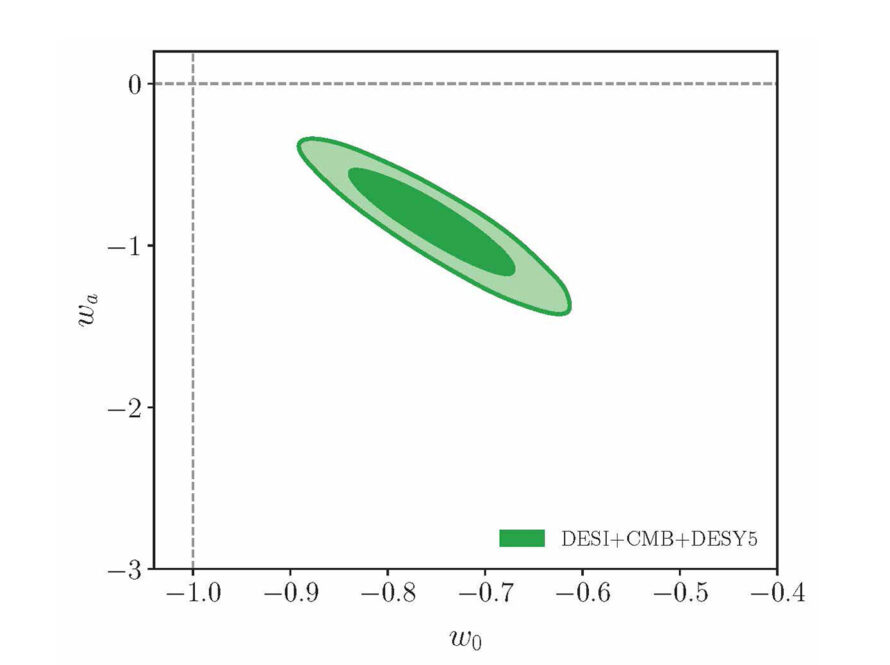}
    \caption{A contour plot in DESI's new dark energy analysis, showing the constraints on parameters of evolving dark energy. (Credit: Cristhian Garcia-Quintero/DESI collaboration)}
    \label{fig:enter-label}
\end{figure}

We have establish that in void-wall cosmology, the expansion rate varies spatially, voids expand faster than the cosmic average, while walls expand slower due to higher matter density. Our local measurements sample a particular region of this cosmic structure, while early Universe measurements average over the entire cosmos. In the early Universe  for redshift $z > 2$, the Universe is more homogeneous, and expansion follows closer to standard $\Lambda$CDM predictions. At intermediate Redshifts $1 < z < 2$, large-scale structures begin forming more prominently. The wall density starts affecting the expansion rate, causing deviations from $\Lambda$CDM. In the late  Universe $z < 1$ our local measurements are affected by our position relative to voids and walls. If we are in or near a void, we would measure a higher local expansion rate.

The effective Hubble parameter in void-wall cosmology can be expressed as:
\be
H_{\text{eff}}(z) = H_{\Lambda\text{CDM}}(z) \cdot [1 + \delta_H(z, \text{position})],
\ee
where $\delta H$ represents the deviation from $\Lambda$CDM due to void-wall effects, which depends on both redshift and position. For the model to resolve the Hubble tension:
\be
\delta_H(z \approx 0, \text{local}) \approx +0.1.
\ee
This creates approximately a $10\%$ increase in the local Hubble constant compared to the global average, matching the observed tension. In the early Universe phase $z > 2$
the Universe is nearly homogeneous and CMB measurements probe this epoch inferring $H_0 = $ 67-68 $km/s/Mpc$ assuming $\Lambda$CDM model evolution. Structure formation begins to grow for redshift $1 < z < 2$. Large-scale structures forms more prominently
and according to DESI 2025 data and other datasets, cosmic acceleration begins weakening here. The wall density effects start modifying the expansion history and the equation of state of dark energy effectively shifts from $w_0\sim -1$ toward less negative values. In the late Universe, $z < 1$ the void-wall structure is fully developed and local measurements sample regions affected by this structure measuring $H_0\approx 73$ km/s/Mpc

The effective dark energy in this model evolves as:
\be
w_{\text{eff}}(z) = w_{\Lambda} + \Delta w(z),
\ee
where $\Delta w(z)$ increases after $z\sim 1-2$, making dark energy less negative and weaker acceleration, consistent with DESI 2025 and other datasets findings. The Friedmann equation becomes:
\be
H^2(z) = H_0^2 \left[\Omega_m(1+z)^3 + \Omega_{\Lambda}\exp\left(3\int_0^z \frac{\Delta w(z')}{1+z'} dz'\right)\right].
\ee
The model predicts directional dependence. The Hubble constant should vary slightly depending on the direction of observation relative to void-wall structures. Measurements at different distance scales should show systematic variations in $H_0$.
A transition in cosmic acceleration around $z\sim 1-2$, occurs as the DESI 2025 data suggests. Modifications to the matter power spectrum at scales corresponding to void-wall structures are expected to occur. The void-wall model naturally explains why cosmic acceleration would weaken at $z\sim  1-2$, as this corresponds to the epoch of significant structure formation. Unlike many other solutions, it does not require introducing new fundamental physics, just accounting for the effects of known structures. The model aligns with the DESI 2025 data,  combined with other datasets, such as weak lensing, supernovae data and CMB data, that cosmic acceleration is weakening in the late Universe.

Detailed numerical simulations would be needed to verify that the magnitude of the effect matches the observed Hubble tension. The model predicts specific patterns in large-scale structure that could be tested with future surveys. The model must remain consistent with other cosmological probes like BAO and supernovae measurements.
The void-wall cosmology model, supported by DESI 2025 data showing weakening acceleration after $\sim 1-2$, provides a physically motivated framework that could potentially resolve the Hubble tension by accounting for the effects of large-scale cosmic structure on expansion history.

The effective cosmic acceleration parameter can be expressed as:
\be
q_{\text{eff}}(z) = f_{\text{v}}(z)q_{\text{v}}(z) 
+ f_{\text{w}}(z)q_{\text{w}}(z),
\ee
where q is the deceleration parameter and $q < 0$ for acceleration. As structure formation progresses, $f_w(z)$ increases for $z < 2$ and $q_w(z)$ is less negative due to less acceleration than $q_v(z)$. Therefore, $q_{\rm eff}(z)$ becomes less negative, indicating weakening acceleration. The DESI 2025 data showing weakening acceleration after $z\sim 1-2$ aligns with this wall-dominated effect, because this redshift range corresponds to when cosmic structure formation becomes significant. The matter distribution shifts from nearly homogeneous to increasingly clustered in walls. The gravitational effects of these walls become more pronounced, counteracting dark energy.

The mechanism can resolve the Hubble tension through a two-fold effect. The overall cosmic acceleration weakens due to wall formation, altering the expansion history from what pure $\Lambda$CDM would predict. Our local measurements of $H_0$ are affected by our position relative to these structures. If we measure within or near voids, we observe higher $H_0$ values. If we measure through walls, we observe lower $H_0$ values.
The net effect depends on our cosmic location and measurement methods.

The wall-dominated mechanism is particularly compelling, because it relies on known physics. It naturally explains the timing of the acceleration weakening $z\sim 1-2$.
It provides a physical basis for apparent dark energy evolution without modifying fundamental physics, and it connects the Hubble tension directly to large-scale structure formation

The Planck Sunyaev-Zeldovich (SZ) catalog, the South Pole Telescope (SPT), and the Atacama Cosmology Telescope (ACT) surveys measure cluster counts as a function of redshift. These datasets can be explored to investigate the void-wall cosmology model. The most relevant SZ-based analyses include: Planck Collaboration especially 2016–2018 releases, cluster number counts and cosmological interpretation. The SPT-SZ and SPTpol surveys studying cluster mass functions and bulk flows. The ACT cluster surveys are combined with X-ray or weak-lensing mass calibrations.

A void–wall scenario that significantly changes late-time expansion locally would manifest as large-scale peculiar velocity flows around structures. Several analyses measurements from the 6dF Galaxy Survey and peculiar velocities of cluster samples tend to prefer coherent flows within $\Lambda$CDM predictions, but the peculiar velocity surveys and cluster velocity measurements yields velocity $200-500~km/sec$, consistent with the velocities of wall filament velocities. Weak-lensing measurements KiDS, DES, HSC and 3D galaxy clustering analyses BOSS, eBOSS, and continuing DESI work constrain the amplitude of matter density fluctuations $\sigma_8$ and the growth of structure. 

\section{Conclusions}

We have presented a comprehensive framework for understanding both the Hubble tension and the apparent evolution of dark energy through the lens of void-wall cosmology. Our analysis demonstrates that properly accounting for the effects of cosmic structure formation can naturally explain these apparent anomalies without requiring modifications to fundamental physics or the introduction of new energy components.
We have shown that the backreaction from cosmic inhomogeneities creates an effective contribution to the cosmic expansion that mimics evolving dark energy when interpreted within homogeneous cosmological models. Our calculations indicate that this effect becomes significant precisely during the epoch when structure formation accelerates $z \sim 1-2$, aligning with the DESI 2025 and other datasets observations of weakening cosmic acceleration at low redshifts. The derived effective equation of state parameter $w_0\approx  -0.76$ emerges naturally from our model without fine-tuning.

We have demonstrated that the differential expansion between void and wall regions creates both directional and scale-dependent variations in the measured Hubble parameter. These variations can account for the of order $9\%$ difference between local measurements $H_0\approx 73~ km/s/Mpc$ and CMB-inferred values $H_0\approx 67-68 ~ km/s/Mpc$. Importantly, our model predicts that measurements probing predominantly void regions or shorter distance scales will yield systematically higher $H_0$ values, consistent with observations from local distance ladder techniques.

We have identified specific observational signatures that can distinguish our void-wall model from competing explanations. These include correlations between the apparent dark energy parameters and large-scale structure, anisotropies in the Hubble flow aligned with the cosmic web, and a characteristic scale-dependence in cosmological parameters that follows the typical void-wall distribution. The preliminary comparison with existing data shows promising agreement, though more detailed observations will be needed for definitive confirmation.

It is proposed that the current cosmological tensions may not indicate a crisis in the standard model but point to the limitations of applying homogeneous approximations to an inhomogeneous Universe. As observational precision continues to improve, properly accounting for cosmic structure becomes essential for accurate cosmological inference.

This perspective also impacts our understanding of dark energy itself. While our model does not eliminate the need for dark energy or a cosmological constant, it demonstrates that its apparent evolution may be an artifact of cosmic structure rather than a fundamental property. This distinction is crucial for theoretical interpretations and future observational strategies.

Several avenues for future work emerge from our analysis. First, more sophisticated numerical simulations incorporating full general relativistic effects would provide more precise quantification of the inhomogeneous terms. Developing observational strategies specifically designed to test the directional and scale-dependent signatures of our model would enable more definitive validation. Our void-wall cosmology model offers a physically motivated, mathematically consistent, and observationally testable resolution to the Hubble tension that aligns with recent evidence for evolving dark energy. By recognizing the significant impact of cosmic structure on our interpretation of cosmological observations, we can potentially resolve apparent tensions without abandoning the standard model's core framework. This approach emphasizes the importance of accounting for cosmic inhomogeneities as we enter an era of precision cosmology, where subtle effects can no longer be neglected in our quest to understand the Universe's evolution.

\section*{Acknowledgments}

Research at the Perimeter Institute for Theoretical Physics is supported by the Government of Canada through industry Canada and by the Province of Ontario through the Ministry of Research and Innovation (MRI).

\end{document}